# Effect of Di-phenyl Siloxane on the Catalytic Activity of Pt on Carbon

**Vijay A. Sethuraman, John W. Weidner, [1]**

Center for Electrochemical Engineering
Department of Chemical Engineering
University of South Carolina, Columbia, SC 29208

**Lesia V. Protsailo**

UTC Power, 195, Governor's Highway
South Windsor, CT 06074

[1] – Corresponding author: Phone: (803) 777-3207; Fax: (803) 777-8265
E-mail: weidner@engr.sc.edu

## Abstract

The effect of silicone on the catalytic activity of Pt for oxygen reduction and hydrogen adsorption was studied using di-phenyl siloxane as a source compound at a rotating disc electrode (RDE). Di-phenyl siloxane did not affect the catalytic activity of Pt when it was injected into the electrolyte. However, it blocked the oxygen reduction reaction when it was premixed with the catalyst. Proton transport was not blocked in either case. We postulate that di-phenyl siloxane induces hydrophobicity and causes local water starvation thereby blocking oxygen transport. Hence, the slow leaching of silicone seals in a fuel cell could cause silicon accumulation in the electrode, which will irreversibly degrade fuel cell performance by blocking oxygen transport to the catalyst sites.

## Introduction

Proton exchange membrane fuel cells (PEMFC) are commonly configured into 'stacks', which comprises of multitude of single-cells arranged together as a single stack. In PEMFC stacks, sealing materials are required to separate the gas compartments from each other, to avoid mixing of the fuel ($H_2$) and the oxidant ($O_2$), and to prevent leakage and loss of fuel. They also serve additional functions such as electrical insulation and control of stack height. Silicone and silicon based elastomers are commonly used as sealing materials in PEMFC stack systems[1] due to their wide operating temperature (-40 to 300 °C), excellent hardness (20 to 60), stress relaxation (up to 25% force retained), good electrical resistivity (> $10^{14}$ Ωcm), and dielectric strength (15-17 kV/mm). Silicones also have excellent functional properties such as good pressure sealing ability (20-200 kPa), low swelling in PEMFC fluids (< 5%), and low permeability to fuel gases and coolants. However, currently used silicone-based seals, gaskets and tubing materials fall short of the U. S. Department of Energy performance targets[2,3]. Degraded or substandard seals might cause fuel leaks and may lead to reduced cell voltage due to mixed potential at the electrodes. In spite of this, there is not much in the literature either on the long-term durability of sealing materials or their effect on the performance at such time-scales. Schulze et al.[4] noted alteration on the color of the membranes where the seal material was in contact after fuel cell operation. Using X-ray photoelectron spectroscopy (XPS), they detected residues of silicone on the surface of the membrane and catalysts. Using scanning electron microscope and energy dispersive X-ray (SEM/EDX) analysis, they observed enrichment of silicone residues on Pt. They speculated that the deposition of Si on the catalyst may change the hydrophobic/hydrophilic characteristics of the electrodes.

In a fuel cell, slow leaching of silicone seals could cause silicone deposition on the electrode. In this work, the effect of such silicon deposition on the catalytic activity of Pt for oxygen reduction and hydrogen adsorption was studied using di-phenyl siloxane $[SiO(C_6H_5)_2]_4$ as a source compound at a rotating disc electrode (RDE).

## Experimental

The effect of di-phenyl siloxane on the catalytic activity of Pt/C was studied by means of RDE experimentation. For the measurements described in the RDE study, Pt/Vulcan XC-72 (20% Pt on Vulcan XC-72R carbon, Johnson Matthey Inc., PA) was used. The electrochemical measurements were conducted at room temperature in a standard electrochemical cell (RDE Cell®, Pine Instrument Company, NC) using a rotating disk electrode setup with a bi-potentiostat (Bi-Stat®, Princeton Applied Research Inc., TN) in conjunction with a rotation-control equipment (Pine Instrument Company, NC). EC-Lab® software (version 8.60, Bio-logic Science Instruments, France) was used to control the bi-potentiostat. Catalyst coated glassy carbon disc electrode (5 mm diameter, 0.1966 $cm^2$ area, DT21 Series, Pine Instrument Company, NC) was scanned between 0 – 1.2 V vs. SHE at a sweep rate of 5 mV/s to represent potentials experienced by an electrode in fuel cell operating conditions. This sweep rate was chosen because lower sweep rates had negligible effects on the oxygen reduction current in the limiting region (0-700 mV vs. SHE). Potentials were determined using a Mercury-mercurous sulfate ($Hg/Hg_2SO_4$) reference electrode. All potentials in this study, however, refer to that of the standard hydrogen electrode (SHE). A high-surface area Pt cylindrical-mesh (5 mm diameter, 50 mm length) attached to a Pt wire (0.5 mm thick, 5 mm length) was used as the counter electrode.

Catalyst coated glassy carbon electrodes were prepared as described by Schmidt et al.[5]. In short, suspensions of 1 mg Pt $ml^{-1}$ were obtained by pulse-sonicating 20 mg Pt/Vulcan catalyst with 15 ml triple-distilled, ultrapure water (Millipore Corporation) in an ice bath (70% duty cycle, 60W, 15 minutes) and 5 ml diluted Nafion solution (5% aqueous solution, 1100 EW; Solution Technology Inc., Mendenhall, PA). Sonication was done using a Braun-Sonic U Type 853973/1 sonicator. Glassy carbon disc served as the substrate for the supported catalyst and was polished to a mirror finish (0.05 μm deagglomerated alumina, Buehler®) prior to catalyst coating. An aliquot of 14 μl catalyst suspension was pipetted onto the carbon substrate, which corresponded to a Pt loading of ~14.1 μg Pt $cm^{-2}$. After evaporation of water for 30 minutes in $N_2$ atmosphere (15 in-Hg, vacuum), the catalyst-Nafion® coated electrode was then immersed in deaerated (UHP Nitrogen, Praxair) 0.5 M Perchloric acid ($HClO_4$, 70%, Ultrapure Reagent Grade, J. T. Baker) for linear sweep voltammetry experiments. All solutions were prepared from ultrapure water (Millipore Inc., 18.2 MΩcm). Following three experiments were performed:

(a) No di-phenyl siloxane was present in the electrolyte. The linear sweep voltammogram obtained from this procedure became the baseline for comparison with those having di-phenyl siloxane in the system.

(b) Di-phenyl siloxane was added to the electrolyte (0.5M HClO$_4$) prior to the ECA and ORR measurements. Two such experiments were performed each with a different concentration of di-phenyl siloxane namely 1µM and 10mM.

(c) In the third case di-phenyl siloxane was premixed with the catalyst such that the dispersion had 10 mM di-phenyl siloxane in it.

## Results and Discussion

Figure 1 shows polarization curves for the oxygen reduction reaction (ORR: $O_2 + 4H^+ + 4e^- \longrightarrow 2H_2O$) on a Pt/Vulcan XC-72R thin-film RDE in 0.5 M HClO$_4$ solution at 25 °C and 1 atm bubbled with O$_2$ with and without di-phenyl siloxane. The ORR current for the case with di-phenyl siloxane (1 µM and 10 mM) in the electrolyte is the same as the no di-phenyl siloxane case. However, the ORR current decreased to very low values when di-phenyl siloxane was premixed in the ink or when di-phenyl siloxane was present in the electrode. Though the data in Figure 1 resulted from a negative sweep, similar trends were observed during the positive sweep as well, (i.e., very low ORR current with di-phenyl siloxane in the electrode).

Figure 2 shows polarization data with no oxygen in the electrolyte. There is no difference in the hydrogen adsorption ($Pt + 2H^+ + 2e^- \longrightarrow Pt-H_2$) peaks in the 25-400 mV potential window for these three cases. The area under this peak is a measure of the electrochemical area (ECA) of Pt and electrochemically contacted Pt surface area through hydrogen adsorption.[6] Therefore the presence of siloxane in the catalyst layer is not blocking active sites. Since there is no difference in the ECA between the electrodes with and without di-phenyl siloxane, proton transport to the catalyst was unaffected by the existence of di-phenyl siloxane either in the electrolyte or in the vicinity of the electrode.

What could possibly explain the behavior where the addition of an impurity blocked oxygen transport to the catalyst surface but had no influence on the proton transport? Figure 3 shows a schematic of the catalyst particle with and without di-phenyl siloxane in the vicinity of the catalyst. Di-phenyl siloxane did not adsorb on the surface of the catalyst within the timescale of this experiment. If it were to adsorb on the surface of the catalyst, then the measured ECA *via* H$_{upd}$ would be lower. Oxygen diffuses through water channels to the catalyst surface whereas protons hop via the sulfonic acid chains in the ionomer as well as through surface diffusion on carbon support. The presence of siloxane in the vicinity of the catalyst blocks oxygen transport through water but proton transport through the ionomer. Since siloxane is hydrophobic, we postulate based on the data that it induces local hydrophobicity when present in the electrode and this hydrophobicity blocks oxygen transport. The ionomer channels, however, remain unaffected. Also, no changes in the CV were observed when the experiment was repeated after several hours, which excludes the possibility of a slow poisoning effect at such time scales where the most dramatic effect on ORR was due to the hydrophobicity caused by the presence of siloxane.

In a fuel cell, degradation of seal materials and their slow leaching could result in their accumulation on the electrode, which will irreversibly degrade fuel cell performance due to the following: (1) silicone induced local hydrophobicity could induce mechanical stresses and (2) local water starvation increases the acidity of the ionomer and accelerates chemical degradation via peroxy-radical attack. Further, the poisoned areas on the electrode will be unavailable to contribute to the fuel cell current. In the cathode, these areas will function as localized hydrogen pumps (under constant current operations) leading to parasitic losses to the power output. In the anode, if hydrogen supply to the catalyst sites is blocked, cell reversal will occur resulting in carbon corrosion on the corresponding area on the cathode[7]. These performance degradation and durability problems can result from any poisoning agent that induces hydrophobicity upon adsorption or deposition.

## Conclusion

Diphenyl-siloxane was used a source compound to understand the poisoning effect of the decomposition products of silicone seals over long-term fuel cell operation. Preliminary RDE data indicated that di-phenyl siloxane, when present in the electrode, blocked oxygen transport but did not hinder proton transport. Since oxygen transport occurs through aqueous media, we postulate that di-phenyl siloxane induced local water starvation upon deposition on catalyst sites which blocked oxygen transport to the catalyst sites. In a fuel cell, local water starvation could cause local mechanical stresses, increase local ionomer acidity and accelerate ionomer degradation via peroxy-radical attack. Local water starvation on the cathode side could block oxygen transport and degrade fuel cell performance.

# Acknowledgement

The US Department of Energy supported this work under contract number DE-FC36-02Al67608, for which the authors are grateful.

# List of Figures

Figure 1: Polarization data (first sweep shown) for the oxygen reduction reaction (ORR) on a Pt/Vulcan XC-72R (14.1 $\mu g_{Pt}$ cm$^{-2}$) thin-film RDE [1200 rpm] in 0.5 M HClO$_4$ solution at 25 °C and 1 atm bubbled with O$_2$ for four different cases: no siloxane in the system (-◇-), 1 µM siloxane (-▲-), 10 mM di-phenyl siloxane (-○-), and di-phenyl siloxane pre-mixed in the electrode (-●-). The electrode was swept in the negative direction at the rate of 5 mV s$^{-1}$. The potential scan was started at 1.0 V vs. SHE and was swept cathodically at the rate of 5 mV s$^{-1}$.

Figure 2: Polarization data on a Pt/Vulcan XC-72R (14.1 $\mu g_{Pt}$ cm$^{-2}$) thin-film RDE [1200 rpm] in deaerated 0.5 M HClO$_4$ solution at 25 °C and 1 atm bubbled with N$_2$ for three different cases: no di-phenyl siloxane in the system (-◇-), 10 mM di-phenyl siloxane (-▲-), and di-phenyl siloxane pre-mixed in the electrode (-○-). The potential scan was started at 1.15 V vs. SHE and was swept cathodically at the rate of 5 mV s$^{-1}$.

Figure 3: Schematic of the supported catalyst with and without diphenyl-siloxane on the catalyst particle.

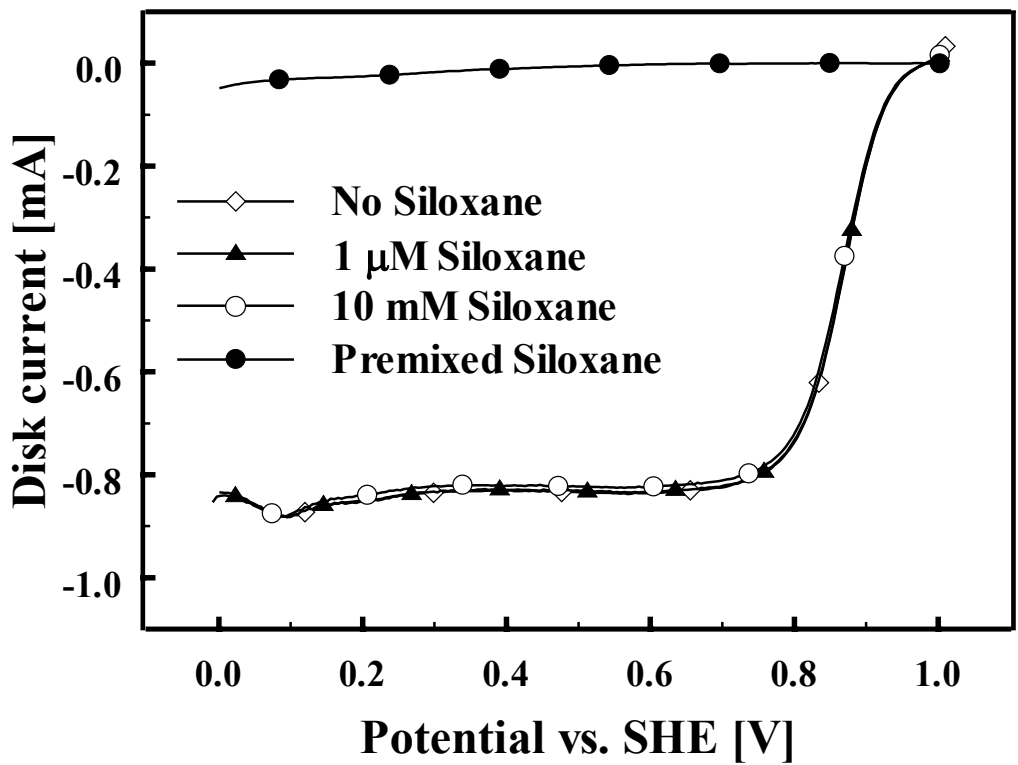

Figure 1: Polarization data (first sweep shown) for the oxygen reduction reaction (ORR) on a Pt/Vulcan XC-72R (14.1 µg Pt cm$^{-2}$) thin-film RDE [1200 rpm] in 0.5 M HClO$_4$ solution at 25 °C and 1 atm bubbled with O$_2$ for four different cases: no siloxane in the system (-◇-), 1 µM siloxane (-▲-), 10 mM siloxane (-o-), and siloxane pre-mixed in the electrode (-●-). The potential scan was started at 1.0 V vs. SHE and was swept cathodically at the rate of 5 mV s$^{-1}$.

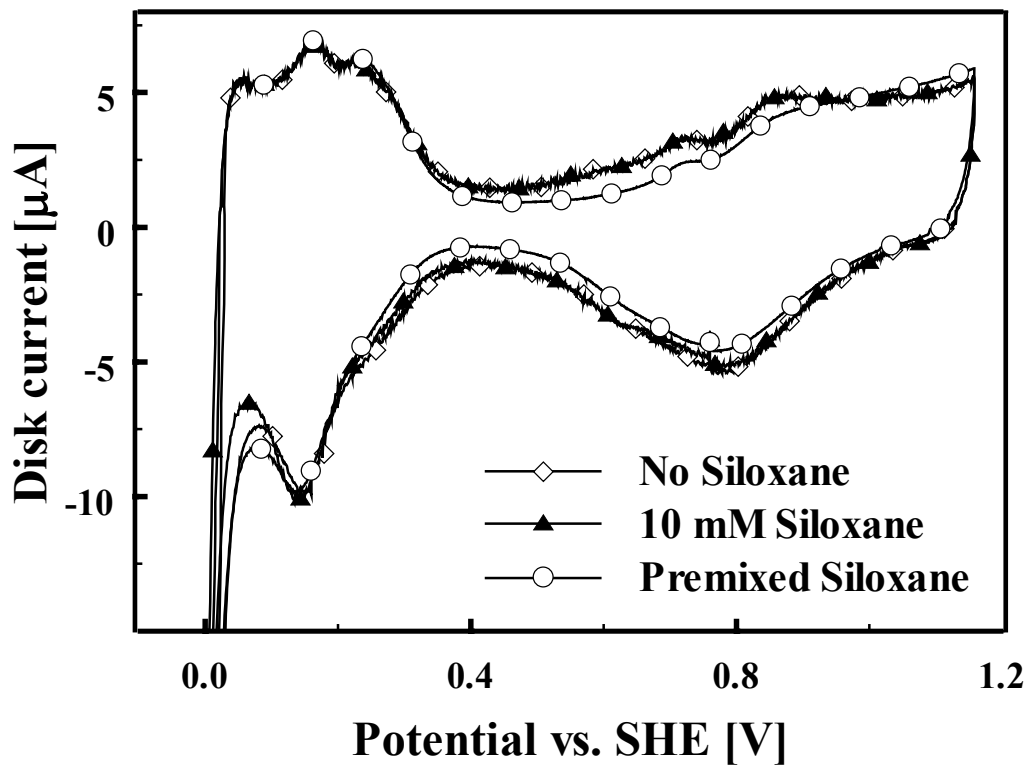

Figure 2: Polarization data on a Pt/Vulcan XC-72R (14.1 μg Pt cm$^{-2}$) thin-film RDE [1200 rpm] in deaerated 0.5 M HClO$_4$ solution at 25 °C and 1 atm bubbled with N$_2$ for three different cases: no siloxane in the system (-◇-), 10 mM siloxane (-▲-), and siloxane pre-mixed in the electrode (-o-). The potential scan was started at 1.15 V vs. SHE and was swept cathodically at the rate of 5 mV s$^{-1}$.

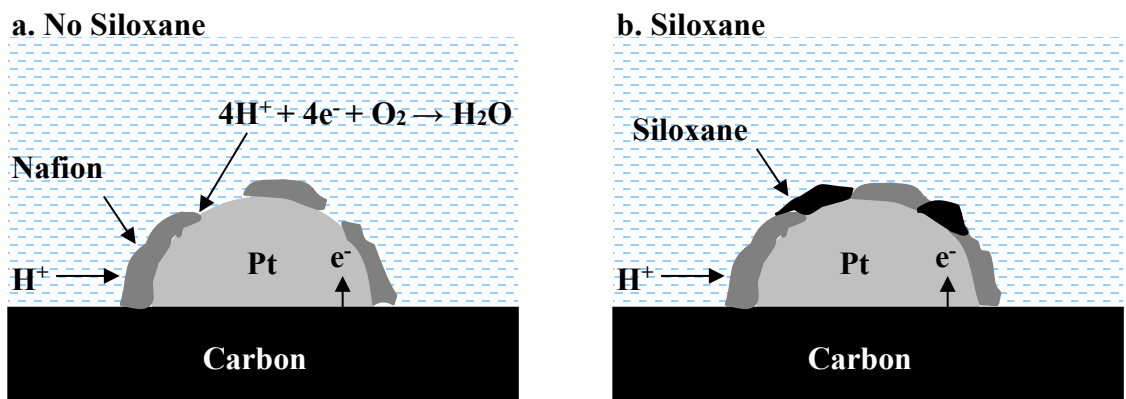

Figure 3: Schematic of the supported catalyst with and without diphenyl-siloxane on the catalyst particle.

# References

1. L. Frisch, *Sealing Technology*, **93**, 7 (2001).

2. T. Clark in *US DoE Hydrogen Program – FY 2004 Progress Report*, p. 493.

3. S. Motupally and T. D. Jarvi, Abstract #1167, 208$^{th}$ Meeting of the Electrochemical Society, October 16-21, Los Angeles, CA.

4. M. Schulze, T. Knöri, A. Schneider, and E. Gülzow, *Journal of Power Sources*, **127**, 222 (2004).

5. T. J. Schmidt, H. A. Gasteiger, G. D. Stäb, P. M. Urban, D. M. Kolb and R. J. Behm, *J. Electrochem. Soc.,* **145**, 2354 (1998).

6 . S. Trasatti and O. A. Petrii, *Pure and Appl. Chem.,* **63**, 711 (1991).

7. C. A. Reiser, L. Bregoli, T. W. Patterson, J. S. Yi, J. D. Yang, M. L. Perry and T. D. Jarvi, *Electrochem. Solid State Letters*, **8**, A273 (2005).